\documentclass[prd,preprintnumbers,eqsecnum,floatfix,a4paper,nofootinbib,twocolumn,superscriptaddress]{revtex4-1}

\usepackage[mathscr]{euscript} 
\usepackage{multirow, newfloat}
\usepackage{color}
\usepackage{calc}
\usepackage{amsmath,amssymb}
\usepackage{bm}
\usepackage{microtype}
\usepackage{booktabs}
\usepackage{times}
\usepackage[varg]{txfonts}
\usepackage[colorlinks, pdfborder={0 0 0}]{hyperref}
\usepackage[utf8]{inputenc}
\definecolor{LinkColor}{rgb}{0.75, 0, 0}
\definecolor{CiteColor}{rgb}{0, 0.5, 0.5}
\definecolor{UrlColor}{rgb}{0, 0, 0.75}
\hypersetup{linkcolor=LinkColor}
\hypersetup{citecolor=CiteColor}
\hypersetup{urlcolor=UrlColor}
\allowdisplaybreaks
\usepackage[caption=false]{subfig}
\usepackage{graphicx}

\usepackage[dvipsnames]{xcolor}
\usepackage{hyperref}
\hypersetup{
  colorlinks=true,
  citecolor=ForestGreen,
  urlcolor= WildStrawberry,
  allbordercolors={0 0 0},
  pdfborderstyle={/S/U/W 1}
}

\def \Hz	{\mathrm{Hz}}

\def \rlist	{\mathscr{U}}

\def \tlist	{\mathscr{T}}

\def \msun	{{M}_\odot}

\def \imrpd {\textsc{IMRPhenomD}}
\def \seobnr {\mathrm{SEOBNRv4ROM}}
 
\def \flow	{f_{\mathrm{low}}}
\def \fref	{f_{\mathrm{ref}}}
\def \fhigh	{f_{\mathrm{high}}}
\def \Mmin	{\zeta}
\def \match {{\mathcal{M}}}
\def \dmax {{\mathcal{D}_{\mathrm{max}}}}
\def \dlambda{{\Delta \lambda}}

\newcommand{\thetaz}{\theta_{0}}
\newcommand{\thetat}{\theta_{3}}
\newcommand{\thetats}{\theta_\mathrm{3S}}

\newcommand{\anstar}[1]{\mathscr{A}_{#1}^{\ast}}

\newcommand{\nth}{\textsuperscript{th}} 

\begin{document}

\title{Effectual template banks for upcoming compact binary searches in Advanced-LIGO Virgo data}

\author{Soumen Roy}		\email{soumen.roy@iitgn.ac.in}
\author{Anand S. Sengupta}	\email{asengupta@iitgn.ac.in}
\affiliation{Indian Institute of Technology Gandhinagar, Gujarat 382355, India.}
\author{Parameswaran Ajith}     	\email{ajith@icts.res.in}
\affiliation{International Centre for Theoretical Sciences, Tata Institute of Fundamental Research, Bangalore 560089, India}
\affiliation{Canadian Institute for Advanced Research, CIFAR Azrieli Global Scholar, MaRS Centre, West Tower, 661 University Ave., Suite 505, Toronto, ON M5G 1M1, Canada}


\begin{abstract}
Recent discoveries of gravitational wave (GW) signals from astrophysical compact binary systems of neutron stars and black holes have firmly established them as prime sources for advanced GW detectors. Theoretical templates of expected signals from such systems have been used to filter the detector data using the \emph{matched filtering} technique. An efficient grid over the parameter space at a fixed minimal match has a direct impact on improving the computational efficiency of these searches. We present the construction of three dimensional template banks (in component masses and an effective spin parameter) by incorporating several new optimizations to the hybrid geometric-random template placement algorithm that we proposed recently. These optimizations allows us to create more efficient template banks in future compact binary searches by shrinking the hybrid banks by $\sim 34\%$ in comparison to the basic algorithm. Such optimised banks are also found to be $\sim 22\%$ 
smaller than the optimized stochastic bank constructed over a nominal range of parameters. We also construct an explicit hybrid template bank with parameters identical to the ``uberbank'' used in the recently-concluded CBC searches in the second observation run of the Advanced LIGO and Virgo detectors. We demonstrate a reduction of more than 53,000 templates over the stochastic template bank at a near-identical coverage as determined by fitting factor studies. A computationally efficient technique for semi-numerical calculation of the parameter space metric, applicable for aligned-spin waveform family, is also outlined. The resulting hybrid template banks can be generated much faster in comparison to the stochastic banks, and are ready to be used in the upcoming observation runs of Advanced LIGO and Virgo.

\end{abstract}


\maketitle 

\section{Introduction}
\label{sec:intro}

Gravitational-wave (GW) observations have opened up a new branch of astronomy with the detection of  signals from several binary black hole (BBH) mergers~\cite{gw150914, gw151226, gw170104, gw170814} in the Advanced LIGO~\cite{advancedLIGO-2015} and Advanced Virgo~\cite{advancedVIRGO-2015} detectors. GW signal from a binary neutron star inspiral~\cite{gw170817} was detected recently, along with several optical counterparts across the electromagnetic spectrum~\cite{gw170817-mma-paper}, heralding the era of multi-messenger astronomy. As the advanced detectors are paced through their planned hardware upgrades, one expects further improvements of their sensitivity and bandwidth. Several advanced detectors such as KAGRA~\cite{ kagra-2013} and LIGO-India~\cite{indigo-proposal-2011} are expected to start operations over the time-scale of a few years. As such, one expects several more detections of GW events to be recorded by this network of advanced detectors. This will usher in a new era of precision GW astronomy. 

GW searches from the inspiral, merger and ringdown phases of compact binary systems are based on two broad techniques: {\emph{modeled}} searches which use theoretical waveforms for such systems as predicted by general relativity and {\emph{unmodelled}} or {\emph{burst}} searches  which assume minimal information about these waveforms. Availability of precise waveforms for these systems allows one to use the \emph{matched filtering} technique to detect weak signals buried in detector noise at a higher statistical significance over burst searches. For example, the GW150914~\cite{gw150914} event was detected by the burst search ~\cite{burst-search} with a significance of $4.1\sigma$ above background whereas, the same event was reported with a signal-to-noise ratio (SNR) of $\sim$24 at a significance $\geq 5.1\sigma$ by a matched-filtering based modeled search~\cite{gw15092014:CBC}. The modeled searches have also played a key role in the detection of the relatively weaker BBH events GW151226~\cite{gw151226}, GW170104~\cite{gw170104} and GW170814~\cite{gw170814}, as well as the binary neutron star (BNS) event GW170817~\cite{gw170817} thus underlining their importance in search pipelines.

Matched-filtering searches involve calculating the correlation between the data and the expected waveform \cite{Dhurandhar-1991, Dhurandhar-1994} and can be shown to be the optimal strategy to search for signals embedded in additive, stationary, Gaussian noise. In reality, however, the detector noise is neither stationary nor Gaussian; therefore one requires additional consistency tests to distinguish between astrophysical signals and noise transients that couple to the detector. Signal consistency tests such as the $\chi^2$-test~\cite{Allen-2005}, trigger coincidence test across multiple detectors \cite{Anand-2008}, etc. help in reducing the false alarm rate and improve the confidence of detections~\cite{ihope-paper} .

Since the signal parameters are not known {\emph{a priori}}, one filters the data using a set of expected signals, each corresponding to a point in the intrinsic parameter space, and are collectively known as the \emph{template bank}. For the first observational run (O1) of Advanced LIGO, the bank was constructed by targeting compact binary systems with individual masses between $(1, 99)\ \msun$, while restricting the total mass up to a maximum of $100 \msun$. The dimensionless aligned-spin magnitude of the individual objects was limited to $0.99$ ~\cite{gw15092014:CBC, lvt151012}. Out of these, objects with mass less than $2\msun$ were considered to be neutron-stars and their  dimensionless spin magnitude was restricted to $0.05$. The template bank consisted of $\sim$250\;000 templates. Template waveforms were modeled using the \textit{effective-one-body} (EOB) formalism~\cite{Taracchini-2014} combined with reduced-order modeling techniques~\cite{Purrer-2014}. In the second observational run (O2) of Advanced LIGO, the template bank was constructed by extending the  total  masses up to $500 \;\msun$ in view of improved low-frequency sensitivity of the detectors~\cite{Canton-2017}. An improved version of aligned-spin EOB waveform model~\cite{Bohe-2017} was used. The aligned-spin magnitudes were similarly limited up to $\sim$0.998. In O2 searches, the number of templates in the bank increased by over $60\%$ in comparison to O1.

The template placement problem for GW searches is one about the construction of a set of points over the parameter space such that for any point chosen at random within its boundaries, one can find a template in the bank within a fiducial distance $\dmax$. The latter is related to the \emph{minimal match} ($\Mmin$)~\cite{Owen-1996} of the bank as: $\dmax = \sqrt{1- \Mmin}$ and is kept fixed for a given template bank. The distance $\dmax$ represents the maximum allowed mismatch between any arbitrary signal in the manifold and the ``nearest'' template in the bank. $\Mmin$ must be carefully chosen as this parameter controls the density of template points. On one hand,  setting a low value can lead to  losses in SNR which in turn can weaken the detection rate. On the other hand, a very high value leads to a dense spacing of templates which in turn can increase the computational cost of the search significantly. Geometrically, the template placement problem is an instance of the \textit{sphere covering problem}~\cite{Prix-2007, Wette-2014, Conway-1999}: where one seeks to cover a given parameter space volume with the smallest set of metric spheres, each of radius $\dmax$. 

The optimum placement of templates in parameter spaces with non-vanishing intrinsic curvature is an open optimization problem under which, one seeks the minimum set of points in the bank at a given fixed, minimal match. Two broad template placement strategies have been developed by the community over the past two decades --- the geometric placement algorithm~\cite{Owen-1999,Babak-2006, Cokelaer-2007} using a quasi-regular lattice of points, and the stochastic construction built from a set of random proposals~\cite{Harry-2009, Babak-2008}. 

The metric on the signal manifold is a crucial input in the geometric approach of template placement. Under this scheme, a set of coordinates is first identified in which 
the parameter-space metric (defined in Section~\ref{sec:waveform}) is almost constant.  In such coordinates, one assumes local flat patches in which a suitably oriented $\anstar{n}$ lattice of points are placed. If the variations are mild, then such local flat patches mesh up neatly to provide adequate coverage over the entire bank. Such banks have been constructed in effective 2-dimensional parameter spaces for both non-spinning and aligned-spin binary neutron stars and neutron star-black hole systems ~\cite{Babak-2006,Cokelaer-2007, Brown-2012, Harry-2014} and have been used in previous LIGO searches \cite{PreviousSearchFirstYear-2009, PreviousSearch-2010, SearchUpperLimitNSBH-2016, SearchGRB150906B}. The most significant drawback of such banks is the amount of fine-tuning required in order to cover ``holes" across local patches because of the mis-alignment of the cells arising from the variation of the metric. Excessive fine-tuning renders the method impractical to be ported across waveform families. It also becomes a challenge to scale the method to higher dimensions. 

The stochastic template placement algorithm, is one where the template bank is built up from a list of random proposals drawn from a uniform distribution over the parameter space. In its original form (which we refer to as bottom-up approach), one starts with an empty bank.  Only those randomly proposed points that lie at a distance greater than $\dmax$ from every existing template in the bank are accepted as new templates. This process continues until the pre-set convergence threshold is reached. The latter is measured from the rate of rejection averaged over a fixed number of acceptances. The distance calculation in the stochastic bank can be sped up if the parameter space metric is available. However, this method can be extended to cases where the metric is not known, in which case the distance is calculated in a brute-force manner by evaluating the overlap integral [see Eq.~\eqref{eq:overlap}] explicitly, at a significant increase in computational cost. Overall, the stochastic method is robust and can be implemented in higher dimensional curved parameter spaces. 

Several recent efforts have been made to combine the space-efficiency of lattice based geometric template placement along with the robustness of stochastic methods.  For example, Capano et al. \cite{Capano-2016} used a seed 4-dimensional geometric bank (constructed over the low-mass range of parameters) as a starting point for the stochastic placement in 4-dimensions. By this approach, they demonstrated a nominal reduction of $\sim 5.5\%$ in the overall bank size over the range of BBH parameters with component masses greater than $2\, \msun$. This bank was used in the  first observing run (O1) of Advanced LIGO. This method was further extended~\cite{Canton-2017} by varying the starting frequency to include the high-mass BBH waveforms and have been used in O2 searches. Other efforts include using binary tree partitions of the parameter space into distinct hyper-rectangles of approximately constant metric over which a geometric or stochastic bank can be placed~\cite{treeBank}.

Along this broad theme of combining the two placement methods, we proposed ~\cite{Roy-2017} a new hybrid geometric-random template placement algorithm and demonstrated its efficacy in a 3-dimensional coordinate system where the parameter-space metric is slowly varying. The idea is the following: at first we start with a large number of random proposals uniformly sprayed over the parameter space. Thereafter, starting from a randomly chosen  point we place suitably oriented $\anstar{3}$ lattice points assuming locally flat patches within the space and remove the sub-set of random proposals that lie within a distance $\dmax$ of the lattice points. The placement is terminated when all the random proposals are exhausted. 
In a head to head comparison with the hybrid bank, we found the vanilla-stochastic bank to have $\sim 28\%$ more templates at an identical coverage for a specific range of search parameters and choice of waveform family. 

In this paper, we present a modified version of the hybrid bank algorithm by incorporating a slew of new optimizations. The new algorithm presented here is suitable for future gravitational wave searches from compact binaries in Advanced-LIGO data. The modifications include the use of a variable lower cut-off frequency and imposing a lower bound on the template duration to extend the mass reach of LIGO searches. We also capitalize on a new degree of freedom by choosing to orientate the $\anstar{3}$ lattices  along a direction governed by the target parameter space boundary, which in turn, has a profound effect on the bank size. We also use numerically computed exact match for removing over coverage of templates in the high-mass region of the parameter space. Each of these optimizations has a profound effect on the overall bank size. The combined effect results in $\sim 34\%$ smaller hybrid banks as compared to the basic algorithm. We also develop an efficient semi-numerical technique for calculating the parameter-space metric that is applicable for a wide range of waveform models. We demonstrate the efficacy of these modifications by explicitly constructing the O2 uber-template bank (using the $\seobnr$ waveform model ~\cite{Bohe-2017} and the power spectrum of the noise from Advanced LIGO's ER10 engineering run) to make a head-to-head comparison against the optimized stochastic template bank used in the recently concluded O2 observation run. 
While both the banks are shown to provide near-identical coverage as determined by fitting factor studies, the stochastic bank is $\sim 15\%$ larger in size. 

The paper is organized as follows: 
In Sec.~\ref{sec:waveform}, we briefly introduce the $\seobnr$ waveform models and some basic terminology used in the paper. We also outline the numerical calculation of the parameter-space metric where relevant partial derivatives  over dimensionless chirp-time coordinates are evaluated using the finite difference method. This simple method is versatile and can be used to calculate the metric for any waveform family. We also show that the contours of constant metric distance are in excellent agreement with those of the numerically calculated ambiguity function.   
In Sec.~\ref{sec:HybridConstruction}, we recapitulate the basic idea of hybrid template-bank construction and highlight the modifications to this algorithm as mentioned earlier.
%
%
In Sec.~\ref{sec:FullBank}, we construct a hybrid bank for the entire search parameter space consisting of BNS, BBH and neutron star black hole (NSBH) systems using the noise PSD from Advanced LIGO's ER10 engineering run\footnote{The ER10 noise PSD was used to generate the optimised stochastic template bank used in LIGO's O2 searches} 
calculated as the harmonic mean of power spectral densities from the LIGO Hanford (H1) and Livingston (L1) detectors with a reference frequency set to $15\, \Hz$. This bank is compared against the \emph{optimized} stochastic bank used for LIGO's CBC searches in O2 data. 
We also present a comparison of the two banks by calculating the signal recovery fraction for $\seobnr$ injections at a canonical signal to noise ratio of 8.
In Sec.~\ref{sec:conclusions}, we summarize the main results of this paper and highlight the suitability of the modified hybrid bank algorithm presented in this paper for future LIGO searches. We also give canonical bank size estimates for the upcoming O3 search expected to start in March 2019.
In Appendix ~\ref{sec:HybridInHigherDimensions}, we present a trivial extension of the basic algorithm for template placement in higher dimensions and calculate theoretical improvements in bank sizes for a few idealized situations. 

\section{Computation of numerical metric for non-precessing black-hole binaries}
\label{sec:waveform}

Recent advances in analytical~\cite{Blanchet-2014, Berti-2006} and numerical~\cite{Pretorius-2005, Campanelli-2006, Baker-2006} relativity has made it possible to construct accurate semi-analytical waveforms describing the entire coalescence of binary black holes, including the orbital inspiral, merger and the subsequent ringdown of the remnant black hole. Waveforms constructed using the \textit{effective-one-body}~\cite{BuonannoEOB-1999, BuonannoEOB-2009, Pan-2011, Tarachini-2012, Tarachini-2014} and phenomenological~\cite{AjithIMR-2007, AjithIMR-2008, Santamaria-2010, AjithIMR-2011,Husa-2016, Khan-2016} approaches have been employed in the searches for GWs from binary black holes, providing significant improvement in the sensitivity of the searches and the accuracy and precision of the estimation of the source parameters. In this paper, we focus on the 
effective-one-body waveform family $\seobnr$. This family of waveforms are faithful in modeling GWs from the inspiral, merger and ringdown of binary black holes with non-precessing spins as established by comparing them against available numerical-relativity simulations over a wide range of the parameters. 

The observed GW signal $h(t; \vec{\lambda}, \vec{\xi})$ from a compact binary system is characterized by a set of intrinsic parameters $\vec{\lambda} \equiv \{\lambda^i\}$ (such as the masses and spin angular momenta of the compact objects) and a set of extrinsic parameters $\vec{\xi} \equiv \{\xi^i\}$ (such as the time of arrival of the signal at the detector and the phase of the signal corresponding to a reference time).
In the case of binaries with non-precessing spins, the intrinsic parameter space is four-dimensional -- $\{m_1, m_2, \chi_1, \chi_2\}$, where $m_1, m_2$ are component masses and $\chi_1, \chi_2$ are the dimensionless spins of the two objects. Spin effects appear as higher order corrections to the inspiral waveform, and the leading spin-dependent term in the post-Newtonian waveform is described by a particular combination of the spins and mass ratio, called the \emph{reduced spin}~\cite{Ajith-2011b}:
\begin{equation}
\chi_r = \chi_s + \delta \, \chi_a - \frac{76 \eta}{113} \chi_s,  
\end{equation}
where $\chi_s$ and $\chi_a$ are symmetric and asymmetric combination of the spins ($[\chi_1 \pm \chi_2]/2$) while $\eta$ and $\delta$ are called symmetric and asymmetric mass ratios: $\eta = m_1 m_2 / (m_1+m_2)^2$ and $\delta = (m_1 - m_2)/(m_1+m_2)$. As a result of this, the dominant spin effects are described by the effective parameter $\chi_r$ and it is possible to mimic, to a good accuracy, the non-precessing waveforms with arbitrary spins by waveforms described by a single effective spin parameter. This essentially means that, to a very good approximation, the effective dimension of the parameter space is three. In this paper, we will be constructing template banks in three intrinsic dimensions and will illustrate that the bank is highly effectual in detecting signals over the full parameter space of non-precessing binaries. 

Matched filtering involves maximizing the following inner product with the data $d(t)$ over the intrinsic and extrinsic parameters 
\begin{equation}
\rho = \max_{\vec \lambda,\, \vec \xi} \left< d ,\: \hat{h}(\vec{\lambda}, \vec{\xi}) \right>,
\end{equation}
where $\hat{h} = {h}/\sqrt{\left<h, h\right>}$ is the normalized waveform and the angular brackets define the following inner product 
\begin{equation}
 \left< a, b \right> := 2 \int_{\flow}^{\fhigh} df \, \frac{ \tilde{a}(f) \, \tilde{b}^*(f) + \tilde{a}^*(f) \, \tilde{b}(f)}{S_h (f)}, 
\label{eq:overlap}
\end{equation}
where $ \tilde{a}(f)$, $\tilde{b}(f)$ denote the Fourier transform of $a(t)$, $b(t)$ respectively; and $S_{h} (f)$ denotes the one-sided detector noise power spectral density. The inner product can be maximized over the extrinsic parameters using efficient numerical algorithms~\cite{Dhurandhar-1994, Owen-1996}, while the maximization over the intrinsic parameters requires the construction of a template bank. In this context, it is useful to define the match as the inner product between two normalized waveforms maximized over the extrinsic parameters:
\begin{equation}
\match (a, b) := \max_{\vec \xi}  \left< \hat a, \hat b \right> .
\end{equation}

Following~\cite{Owen-1996}, the match between two nearby waveforms, whose intrinsic parameters differ by $\Delta \vec \lambda$, can be Taylor expanded upto the quadratic terms which in turn, can be rearranged to express the \emph{mismatch} $(1-\match)$ in terms of the parameter space metric $g_{ij}$ as 
\begin{equation}
1 - \match \simeq g_{ij} \, \dlambda^i \, \dlambda^j, 
\label{eq:MisMatch}
\end{equation}
where the metric is given by
\begin{equation}
 g_{ij} := -\frac{1}{2}  \left . \frac{\partial^2 \mathcal{M}}{\partial \dlambda^i \: \partial \dlambda^j} \right \vert_{\dlambda = 0}.
\end{equation}
We compute the metric by evaluating the Fisher information matrix~\cite{Owen-1999} of the waveforms over the full parameter space and then projecting out the dimensions corresponding to the extrinsic parameters~\cite{Buonanno-2003}. The elements of the Fisher matrix are given by 
\begin{equation}
\tilde \Gamma_{\alpha\beta} = \left< \partial_\alpha \hat{h}, \partial_\beta \hat{h} \right>,
\end{equation}

\begin{figure*}
\centering
    \includegraphics[width=0.90\textwidth]{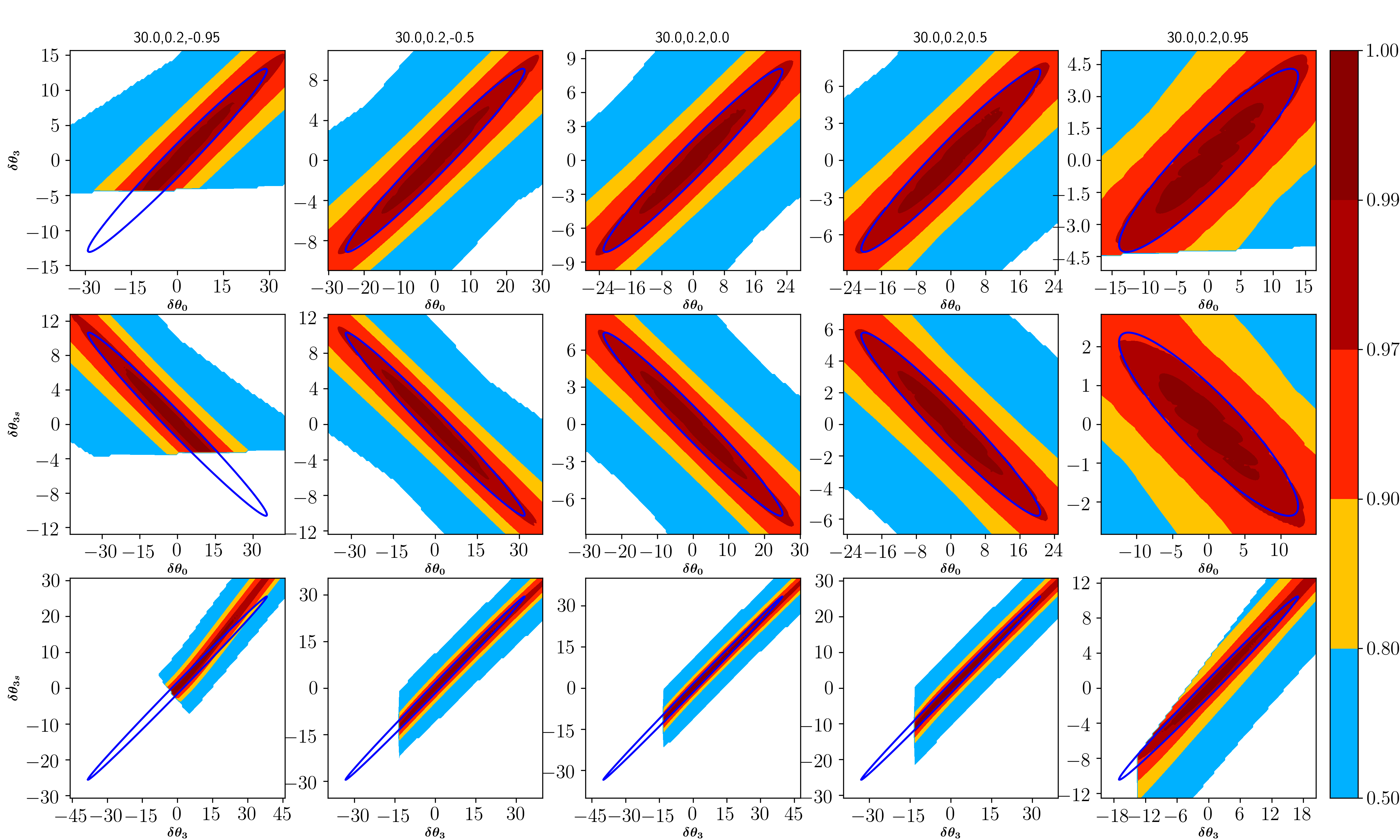}
  \caption{Comparison between the $\match= 0.97$ ellipses computed from the resulting semi-numerical metric (blue ellipses) and the contours of the
numerically calculated match function, for the non-precessing $\seobnr$ waveform model. The  rows correspond to different orthogonal slices over the $\{\thetaz, \thetat, \thetats \}$ coordinate system. The columns correspond to different values of the dimensionless equal-aligned-spin parameters $\chi_1 = \chi_2 = \left[-0.95, -0.5, 0.0, 0.5, 0.95\right]$. The total mass is chosen to be $M = 30\, \msun$ with symmetric mass ratio $\eta = 0.2$. All computations assume ER10 noise spectral density which is measured from the harmonic average over H1 and L1 noise power spectral densities with a lower cutoff frequency $15\, \Hz$. Some of the shaded areas are cut-off along the boundary of the physical parameter space.
}
\label{fig:metric_exact_mismatch_comparison}
\end{figure*}

where $\partial_\alpha$ denotes the partial derivative of the waveform with respect to the $\alpha$\nth parameter (this includes intrinsic and extrinsic parameters). The derivatives are computed numerically using finite difference methods. 
In particular, we compute the derivatives of the amplitude $A$ and phase $\Psi$ of the frequency domain waveforms and compute the Fisher matrix from these derivatives. 

\begin{equation}
  \tilde \Gamma_{\alpha\beta} \simeq  \frac{1}{2||h||^2}\left[ \left< A\partial_\alpha \Psi, A\partial_\beta \Psi \right> + \left< \partial_\alpha A, \partial_\beta A\right>  \right]
  \label{eq:FisherMatrix}
\end{equation}

The Fisher matrix is computed in the 5-dimensional parameter space of three intrinsic parameters ($\vec{\lambda}$) describing the total mass ($M$), symmetric mass ratio ($\eta$) and a reduced spin parameter ($\chi_{r}$) and two extrinsic parameters describing the time of coalescence $t_0$ and phase $\varphi_0$ at coalescence. Wherever possible, the numerical derivatives are carried out by using the first-order accurate central finite differencing scheme. At the boundaries of the parameter space, where the central difference scheme cannot be applied, we use the first-order accurate forward/backward schemes as appropriate. 
The inner product integrals in the expression of the Fischer matrix Eq.~\eqref{eq:FisherMatrix} are evaluated numerically. 

We describe the intrinsic parameters in terms of three dimensionless chirp times $\thetaz, \thetat$ and $\thetats$~\cite{Ajith-2014}, which are related to the total mass, symmetric mass ratio and the \emph{reduced} spin parameters in the following way: 
\begin{eqnarray}
M  & = & \frac{5}{32 \, \pi^2 f_0}\frac{\thetat}{\thetaz}, \nonumber \\
\eta & = & \left(\frac{16 \, \pi^5}{25} \frac{\thetaz^2}{\thetat^5} \right)^{1/3}, \nonumber \\
\chi_r & = & \frac{48 \, \pi \, \thetats}{113 \, \thetat},
\label{eq:MassSpinToThetas}
\end{eqnarray}
where $f_0$ is the instantaneous frequency of the chirp waveform at the fiducial starting time.
The choice of this set of parameters was motivated by the fact that, when expressed in these coordinates, the metric is found to be slowly varying over the parameter space. A non-singular Jacobian transformation $\mathbf{J}$ is used to transform the metric $\mathbf{\tilde \Gamma}$ over the $\{M, \eta, \chi_r, t_0, \varphi_0\}$ space  to the metric $\mathbf{\Gamma}$ over $\{\thetaz, \thetat, \thetats, t_0, \varphi_0\}$ space:
\begin{equation}
  \mathbf{\Gamma} = \mathbf{J} ~ \mathbf{\tilde \Gamma} ~ \mathbf{J}^{\mathrm{T}}.
\end{equation}

Finally, we compute the 3-dimensional metric $\mathbf{g}$ over the intrinsic parameters $\{\thetaz, \thetat, \thetats\}$ by projecting out $\mathbf{\Gamma}$ along directions that are orthogonal to $t_0$ and $\varphi_0$. Operationally, this is equivalent to evaluating the Schur complement of the extrinsic parameters block ($\mathbf{\Gamma}_{4}$), explicitly given by:
 \begin{equation}
   \mathbf{g} = \mathbf{\Gamma} / \mathbf{\Gamma}_{4} \coloneqq \mathbf{\Gamma}_1 - \mathbf{\Gamma}_2^{\mathrm{T}} \, \mathbf{\Gamma}_{4}^{-1} \, \mathbf{\Gamma}_2,
 \end{equation}
where $\mathbf{\Gamma}_1$ is the intrinsic parameters block,  and $\mathbf{\Gamma}_2$ refers to the block of cross-terms between intrinsic and extrinsic parameters of $\mathbf{\Gamma}$. 

We have outlined the method to compute the metric $\mathbf{g}$ over 3-dimensional intrinsic parameters $\vec{\theta} \equiv \{\thetaz, \thetat, \thetats\}$ as given by Eq.~\eqref{eq:MassSpinToThetas}, where the two spin parameters $\chi_{1,2}$ have been condensed to a single effective parameter $\chi_r$. It is obvious that $\chi_r \mapsto (\chi_1, \chi_2)$ is not bijective. A one-to-one, onto mapping between the reduced spin and individual spin parameters is imposed by assuming that the BBH and BNS systems have equal-aligned spins ($\chi_1=\chi_2$), and that the neutron star companions of NSBH systems are non-spinning. The metric for aligned-spin NSBH system with non-spinning neutron star component is found to have a good agreement with the spinning neutron star case up to dimensionless aligned-spin magnitude $\leq 0.1 $. 
 
Figure~\ref{fig:metric_exact_mismatch_comparison} shows the comparison between mismatch contours (solid blue ellipses) of $\mathbf{g}$ (calculated using Eq.~\eqref{eq:MisMatch}) and the contours of
the exact numerical match function (Eq.~\ref{eq:overlap}). From the two dimensional slices over various planes, it is evident that there is an excellent agreement between the two, thereby validating the method outlined in this section. We have also successfully tested this method to calculate the metric for  $\imrpd$ and TaylorF2 waveform families.


\section{Optimizing the hybrid bank algorithm}
\label{sec:HybridConstruction}

The {\texttt{lalapps\_cbc\_sbank}} code is a straightforward vanilla implementation of the stochastic template placement algorithm~\cite{Harry-2009} available in the {\texttt{LALSuite}}~\cite{lalsuite} code base. This code was used as a benchmark to measure the comparative efficiency of the hybrid algorithm as presented in our previous work ~\cite{Roy-2017}. It was shown that the hybrid banks were about $21\%$ smaller as compared to the vanilla stochastic banks. Recently, new optimizations to the stochastic algorithm were introduced in a paper by Del Canton et al.~\cite{Canton-2017} in the context of constructing template banks for LIGO's O2 searches, so as to increase the mass-reach of the search beyond O1. 
Our present work is largely motivated to incorporate similar optimizations within the hybrid bank algorithm and also explore other improvements to further reduce the bank size. In Table~\ref{tab:O2bankSummary}, we show that these improvements finally lead to a hybrid bank for O2 search that has $53\, 000$ fewer templates than the optimized stochastic bank at an identical coverage. With these changes in place, we aim to provide a viable template bank with the smallest footprint for upcoming LIGO searches. We also find that these changes lead to huge computational advantages, such that the hybrid banks can be created much faster as compared to the optimized stochastic banks.
   
\begin{figure}
\centering
  \includegraphics[width=0.475\textwidth]{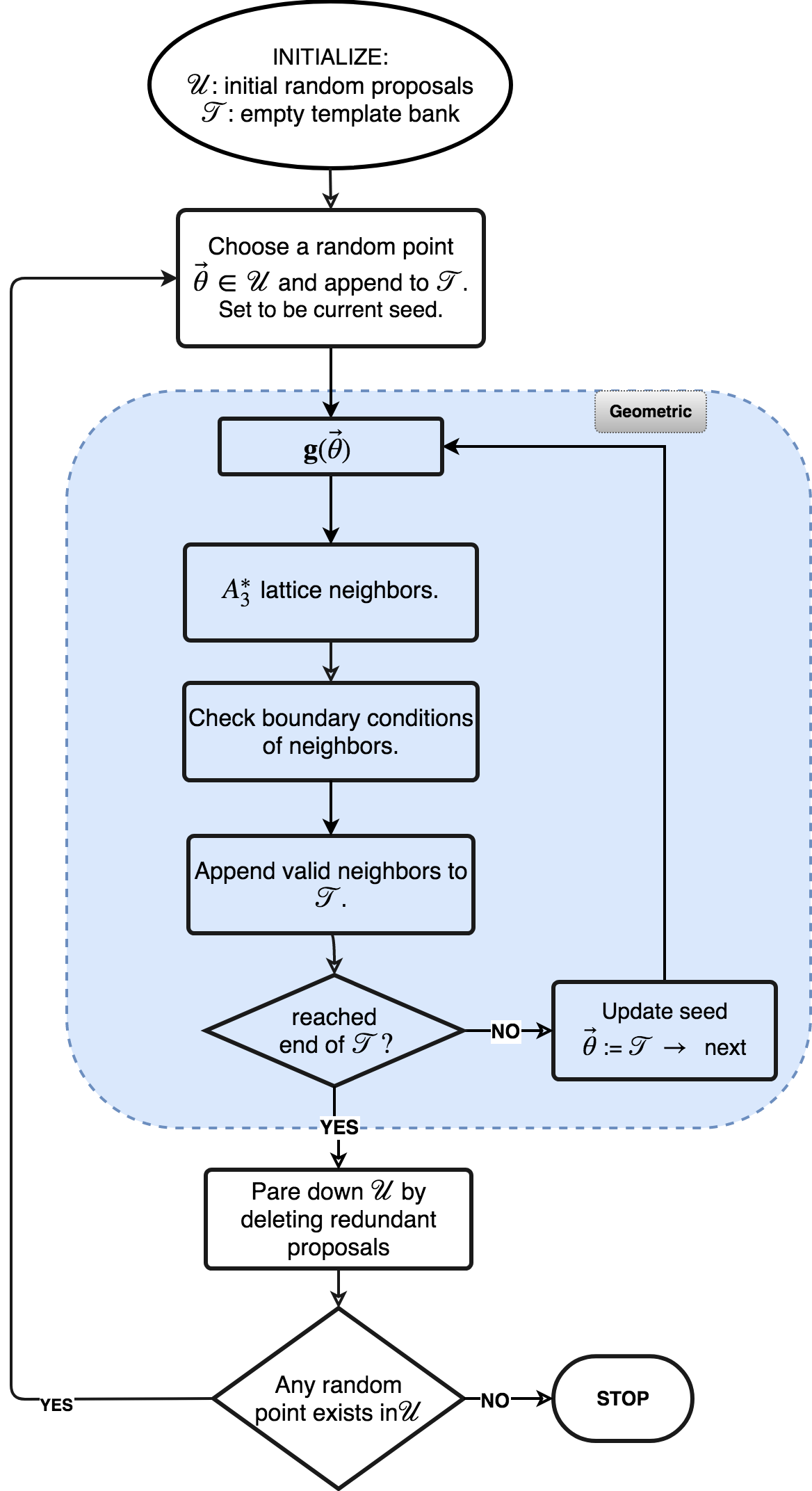}
  \caption{Flowchart of the hybrid template placement algorithm. $\rlist$ and $\tlist$ are the lists of initial random proposals and template points respectively. See the text in Section~\ref{sec:HybridConstruction} for a complete description.}
  \label{fig:Algorithm}
\end{figure}

We start this section by recalling the classic geometric-random hybrid template placement algorithm as shown in Fig.~\ref{fig:Algorithm}. We then explain each of the new optimizations and indicate the effect of these improvements on the final bank size.

As mentioned in Section~\ref{sec:waveform}, the templates are placed in the dimensionless chirp time coordinates $\vec \theta$ over which the parameter space metric varies slowly. The hybrid algorithm starts by initializing a large set of random proposals $\rlist$, uniformly distributed over $\vec \theta$. After this, a random initial seed is chosen from $\rlist$ and appended to an empty template list $\tlist$. Thereafter, it enters the geometric part of the algorithm (refer to the shaded part of Fig.~\ref{fig:Algorithm}) where, the metric $\mathbf{g}(\vec \theta)$ at the seed point is evaluated at first using which, a locally flat coordinate patch (assumed to be valid upto a distance $2\mathcal{D}_{\mathrm{max}}$ around the centre) is created. The positions of the local $\anstar{3}$ lattice neighbors are calculated with respect to this local coordinate frame using a computationally efficient method outlined in Appendix~\ref{app:AnstarNeighbor}. Finally, they are transformed back to the target dimensionless chirp-time coordinates using the procedure outlined below: scaling and rotation matrices are constructed using the eigenvalues $\{\Lambda_{i}\}$ and eigenvectors $\{\vec{e}_{i}\}$ (where $i=1,2,3$) of the metric $\mathbf{g}(\vec \theta)$ in the following manner: the columns of the rotation matrix $\mathbf{\mathcal{R}}$ consist of the eigenvectors such that the element $\mathcal{R}_{ij}$ is the $j^{\rm th}$ element of the $i^{\rm th}$ eigenvector. The scaling matrix $\mathbf{\mathcal{S}}$ is diagonal with elements $\mathcal{S}_{ii} = \dmax/\sqrt{\Lambda_{i}}$. Let $\bar N^p_k$ be the $k$\nth coordinate of the $p$\nth $\anstar{3}$ neighbour (where $k=1,2,3$ and $p=1,2,\ldots,14$) in the 3-dimensional local flat coordinate system $(\hat e_1, \hat e_2, \hat e_3)$. The transformation which converts them to the $\vec \theta$ coordinates is given by:
\begin{equation}
\label{eq:LocalNeighbor}
 N_{i}^p = \sum_{j,k=1}^3 \mathcal{R}_{ij}^{T} \, \mathcal{S}_{jk} \, \bar N_{k}^p.
\end{equation}


Out of the 14 $\anstar{3}$ neighbors, the ones that lie within the search parameter space boundaries and are spaced farther than a distance $\dmax$ from all other templates in $\tlist$ are appended to the template list. Thereafter the next point in $\tlist$ is re-assigned as the current seed point which leads to the next geometric iteration.

The geometrical lattice of templates grows iteratively  until no new points can be appended any further to $\tlist$. This situation happens most commonly near the vicinity of the parameter space boundary and also in extremely narrow regions of the parameter space where the effective dimension is less than that of the search space. In such cases, the $\anstar{3}$ neighbors, when projected back to the target space, lie outside the physical parameter space thereby halting the geometric iterations. The latter may also be interrupted in regions of the parameter space where the metric varies rapidly. In such cases, one expects a small relative change of orientation of the eigenvectors of $\mathbf{{g}}$ even for nearby points as a result of which, the local flat patches do not mesh properly. This results in an imperfect $\anstar{3}$ lattice with 'holes' in the bank. In a flat parameter space (where the metric is constant everywhere), the hybrid construction will result in a perfect $\anstar{3}$ lattice in the bulk of the search volume, excluding the vicinity of the boundary. 

Once the iterative geometric placement is over, random proposals from $\rlist$ that lie within a distance $\dmax$ from the templates are eliminated. The geometric process is restarted by seeding a new random proposal from the remaining points in $\rlist$. This process continues until $\rlist$ is completely exhausted. The final bank size weakly depends ($\sim$1\% variability) on the choice of the initial seed. By first spraying a set of random proposals over the physical parameter space and later eliminating the ones that lie  inside the minimal-match ``spheres'' of the accepted templates, we make sure that any hole left behind due to varying curvature automatically get covered by the remaining random proposals that aren't swept-up by the deletion criteria mentioned above.

We now present new optimizations to the classic hybrid template placement algorithm.

\subsection{Orientation of the $\anstar{3}$ lattice} 

Once a local flat patch $\{ \hat e_1, \hat e_2, \hat e_3 \}$ is constructed, we still have the freedom to rotate these coordinates by three Euler angles $\alpha$, $\beta$ and $\gamma$. Such a rotation affects the alignment of the symmetry axis of the $\anstar{3}$ lattice that is placed in the local patch. Due to the finite volume and boundary effects, it is reasonable to assume that a preferred axis of rotation exists such that the largest fraction of geometrical lattice neighbours lie within the parameter space boundaries. 
By extension, if there is no boundary (infinite volume), then there is no such preferred axis and thus there is no need to orientate the $\anstar{3}$ lattice within the local flat patch. 


We now outline a Monte Carlo simulation to find the appropriate set of Euler angles. First, we generate a set of $n_0$ random points $U_{r}$ uniformly distributed within the range of search parameters in dimensionless chirp time coordinates.
For each of these points, the metric $\mathbf{g}$ is calculated, using which a local, flat coordinate system $\{ \hat e_1, \hat e_2, \hat e_3 \}$ is set up. 
As mentioned earlier, these axes are not unique: one can always rotate them about any axis passing through the origin. For every such rotation, the local coordinates of the $\anstar{3}$ lattice neighbors change, which in turn affect their coordinates with respect to the chirp-time coordinates. The aim is in finding the rotation for which a maximum number of neighbors lie  inside the parameter space boundaries.

For each point in $U_r$, we consider three successive rotation angles $\alpha$, $\beta$ and $\gamma$ about $\hat e_1$, $\hat e_2$ and $\hat e_3$-axis respectively. The coordinates of the neighbors in the $\vec \theta$ space can be calculated by modifying Eq.~\eqref{eq:LocalNeighbor} as:
\begin{equation}
   {N}_{i}^q = \sum_{j,k,l=1}^3 \mathcal{R}_{ij}^{T} \, \mathcal{S}_{jk} \, \mathcal{E}_{kl}(\alpha, \beta, \gamma) \, \bar N_{l}^q,
\end{equation}
where $\mathcal{E}(\alpha, \beta, \gamma)$ is the multiplication of the three successive rotation matrices $\mathcal{E}_{\hat e_1}(\alpha)$, $\mathcal{E}_{\hat e_2}(\beta)$ and $\mathcal{E}_{\hat e_3}(\gamma)$ about the $\hat e_1$, $\hat e_2$ and $\hat e_3$-axis respectively. Suppose $\mathscr{Z}_{p}$ $\anstar{3}$ lattice neighbors (out of 14) of the $p$\nth  random point of $U_{r}$ lie inside the parameter space, then the average number of valid neighbors can be defined as
\begin{equation}
  \left \langle \, \mathscr{Z}(\alpha, \beta, \gamma) \, \right \rangle = \frac{1}{14 \, n_0}\sum_{p=1}^{n_0}\mathscr{Z}_p(\alpha, \beta, \gamma).
\end{equation}

To find the optimum set of angles, we maximize the quantity $\left \langle \mathscr{Z}(\alpha, \beta, \gamma) \right \rangle$ over $\alpha$, $\beta$ and $\gamma$. This leads to the biggest geometrical lattice of templates in each of the geometric iterations of the hybrid algorithm.

We have exploited this freedom of orientating the local flat orthogonal coordinates to construct the hybrid O2 uber bank by averaging over $n_0 = 10\, 000$ random points. It is to be noted that this optimization depends on the overall volume and boundaries of the target search space. The optimally oriented hybrid O2 uber bank has $\sim$ 10\% fewer templates in comparison to the one where such optimal rotations were not applied. 


\subsection{Incorporation of the exact match function}

In the hybrid template placement algorithm, the metric approximation is used in two different ways: (a) to find the local $\anstar{3}$ lattice neighbors of a point and (b) to calculate the distance between any two fiducial points in the parameter space. 
It turns out that the semi-numerical metric approximation becomes inaccurate at high mass regions (especially in the BBH region of the search space with total mass $M>40\, \msun$). In such regions, the mismatch contours calculated in Eq.~\eqref{eq:MisMatch} using the semi-numerical metric is smaller than the contours of the exact match function (Eq.~\eqref{eq:overlap} ). This affects both the geometric placement as well as the elimination of the random proposals from $\rlist$ and leads to an over-coverage in the resulting bank. To get around this problem, we cannot fully rely on the semi-numerical metric and have to use the exact match function.


However, the inclusion of the exact match function (in favour of the more economical metric distance) in the bank construction may lead to computational inefficiencies when comparing the distance of a fiducial point from all the templates in a very large list $\tlist$. As such, we refine the template placement part of the algorithm in two steps: at first, we create a bounding box around the fiducial point using the semi-numerical metric and scale it up by a factor $\mu \geq 1$. 
The dimensions of the bounding box~\cite{Tavian-2014} along $\vec e_i$ are given in terms of the elements of the rotation matrix $\mathcal{R}$ and the eigenvalues of $\mathbf{g}$ as: 
\begin{equation}
	\Delta \theta_i = 2 \dmax \sqrt{\mu \sum_{j=1}^3 \frac{\mathcal{R}_{ij}^2}{\Lambda_j}},	
\end{equation}
which is in agreement with the expressions found in ~\cite{Anand-2008}.


Thereafter the templates in $\tlist$ that lie within this box are identified. This is further refined by retaining only those template points whose metric distance from the fiducial point are found to be less than or equal to $\sqrt{\mu} \dmax$. After this point we are left with a much smaller list of template points against which an exact match is calculated for determining valid $\anstar{3}$ neighbours (required in the geometric iteration) and also in the elimination of initial proposals from $\rlist$. The scaling factor $\mu$ can be treated as a tunable parameter. In our experiments with O2 uber bank construction, we find that $\mu = 3$ for low mass systems and $\mu=5$ for systems with total mass greater than $40\, \msun$ works adequately well.




We have implemented the exact match function within the hybrid bank placement algorithm and tested it by constructing the O2 hybrid uber bank. By a suitable choice of the scale factor $\mu$ we found that the resulting bank had $\sim 9\%$ fewer templates, as compared to a hybrid bank constructed using only the semi-numerical metric. 

\subsection{Inclusion of an optimal starting frequency}

The vanilla stochastic algorithm was spruced up by the inclusion of a (variable) optimal starting frequency at a maximum allowed (small) SNR loss by Del Canton et al.\cite{Canton-2017}  for constructing the template bank used in LIGO's O2 CBC searches. This modification led to a bank which could be used to search for high mass BBH systems in the data. For a given binary system (specified by its intrinsic parameters $\vec \lambda$), this optimum lower cut-off frequency is found by solving the equation that corresponds to a fixed loss in signal power: $R(\flow)=R^{\ast}$ where $R^{\ast}$ is taken to be $0.995$ corresponding to a nominal loss of SNR of 0.5\% at a reference lower-cutoff frequency set to $\fref = 15 Hz$ below which there is no appreciable signal power for any template waveform and where, 
\begin{equation}
  R(\flow) = \int_{\flow}^{\fhigh} \frac{\tilde{A}^{2}{(f; \vec \lambda)}}{S_{h}(f)}\, df \bigg/ \int_{\fref}^{\fhigh} \frac{\tilde{A}^{2}{(f; \vec \lambda)}}{S_{h}(f)}\, df .
  \label{eq:VaribleFlow}
\end{equation}
In the above equation, $\tilde{A}(f; \vec \lambda)$ is the amplitude of the template waveform in the frequency domain. On the flip side, the inclusion of very high-mass BBH systems in the template bank also increases the rate of triggers and it becomes difficult to use the $\chi^2$ signal consistency test. This is avoided to some extent by incorporating a lower-bound on the filter waveform length to exclude very short-duration templates. Such a bound is chosen empirically from observation of the background distribution. 

As the variable $\flow$ also has a profound effect on the size of the template metric ellipses which in turn affect the hybrid template placement algorithm, we have also incorporated this feature as explained below.



We begin by first pointing out that using the optimized lower cutoff frequency is quite straight-forward in the stochastic placement algorithm as it is used only for evaluating the exact match integrals. On the other hand, for the hybrid placement, the optimized lower cutoff frequency needs to be incorporated to calculate the metric $\mathbf{g}$. As outlined in the previous section, this involves transformations back and forth between the physical parameters and the dimensionless chirp time parameters. As such, some care is needed in including this feature. We modify the hybrid algorithm in the following way: first, we initialize the set of random proposals $\rlist$ over the $\vec \theta$ space by assuming a fixed starting frequency $f_0  = \fref$ in the Eq.~\eqref{eq:MassSpinToThetas}. Second, we use an optimized lower cutoff frequency to evaluate the inner product integrals of the Fisher matrix~(Eq.~\ref{eq:FisherMatrix}) elements over the physical parameter space $\vec{\lambda}\equiv \{M, \eta, \chi_r, t_0, \varphi_0\}$. Finally the Jacobian transformation matrix is evaluated assuming a fixed $f_0 = \fref$. Such a protocol enables us to calculate $\mathbf{g}$ with an optimal $\flow$ which in turn affects the co-ordinates of the $\anstar{3}$ lattice points. The fixed reference frequency $\fref$ is also used for determining the proposal elimination criteria where proposals in $\rlist$ are removed if they lie within the minimal match distance from existing  templates in the $\tlist$. Inclusion of this feature has a significant effect in reducing the overall hybrid bank size. 

\begin{table}[htbp!]
\centering
\begin{tabular}{l | l r r }
\toprule[1pt]
\toprule[1pt]
Type of optimization &  \multicolumn{2}{c}{Bank size}  &  \\
\cmidrule[0.8pt]{2-3}
 & Stochastic & Hybrid & \\
\midrule[1pt]
 Vanilla & 25\,208 &   23\,763 &   \\ 
 \ \ + Inclusion of exact match &  & 19\,628 \\
\ \ + Orientation of $\anstar{3}$ lattice &  &18\,078 \\
\ \ + Variable starting frequency & 20\,070  & 15\,671  \\
\bottomrule[1pt]
\bottomrule[1pt]
\end{tabular}
\caption{Comparison of the template bank sizes for the hybrid template placement algorithm using a combination of new optimizations presented in Section~\ref{sec:HybridConstruction} against the vanilla/optimized stochastic method. Their combined effect results in $\sim 34\%$ smaller hybrid banks as compared to the basic algorithm~\cite{Roy-2017} at the cost of $0.5\%$ loss in SNR. }
\label{tab:BankEfficiency}
\end{table}

\begin{figure*}[htbp]
\centering
\subfloat[O2 stochastic-uber bank]{\label{fig:O2_uber_bank}%
    			{\includegraphics[width=0.495\textwidth]{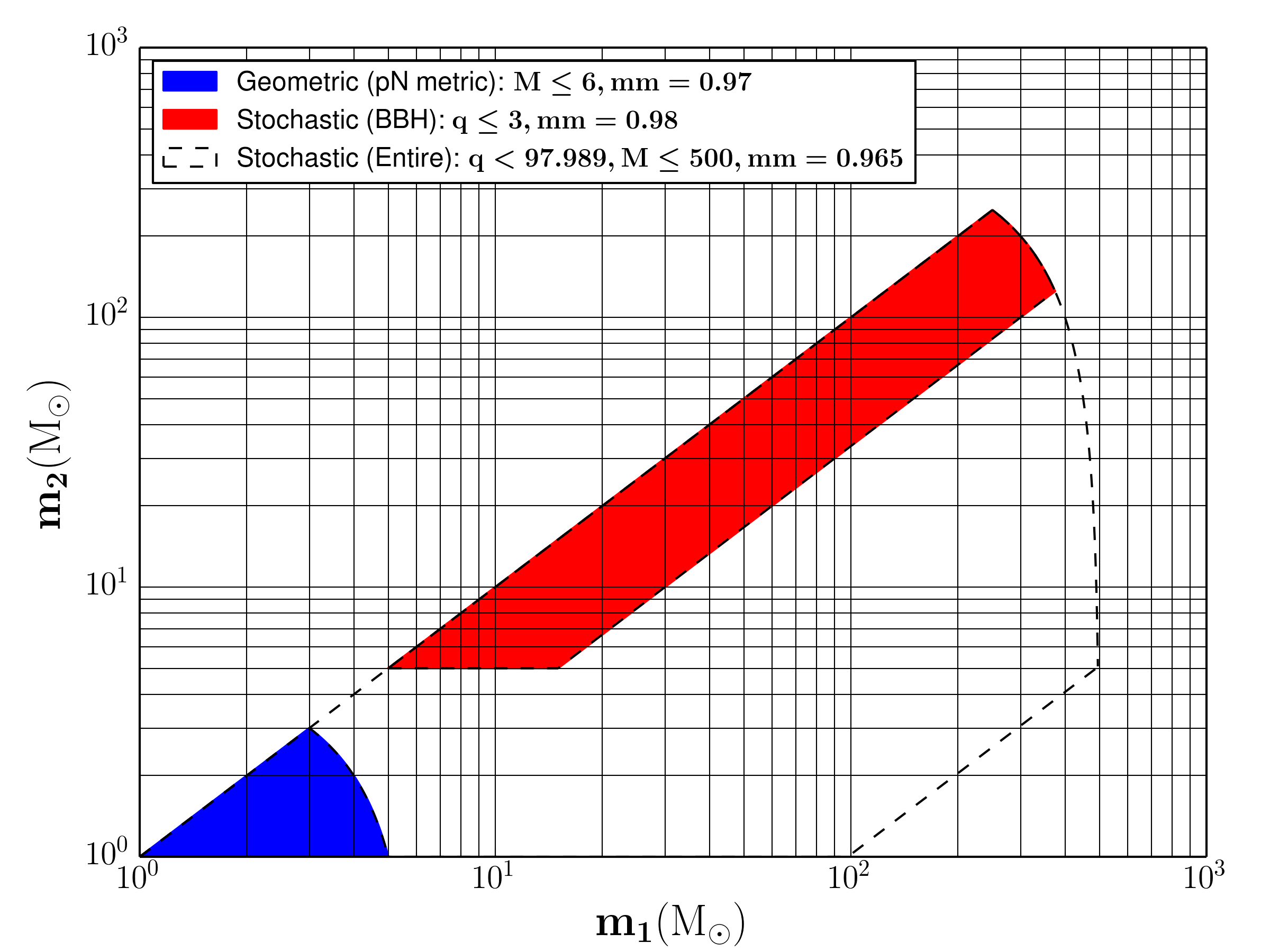} }}%
    \subfloat[O2 hybrid bank]{\label{fig:O2_hybrid_bank}%
    			{\includegraphics[width=0.495\textwidth]{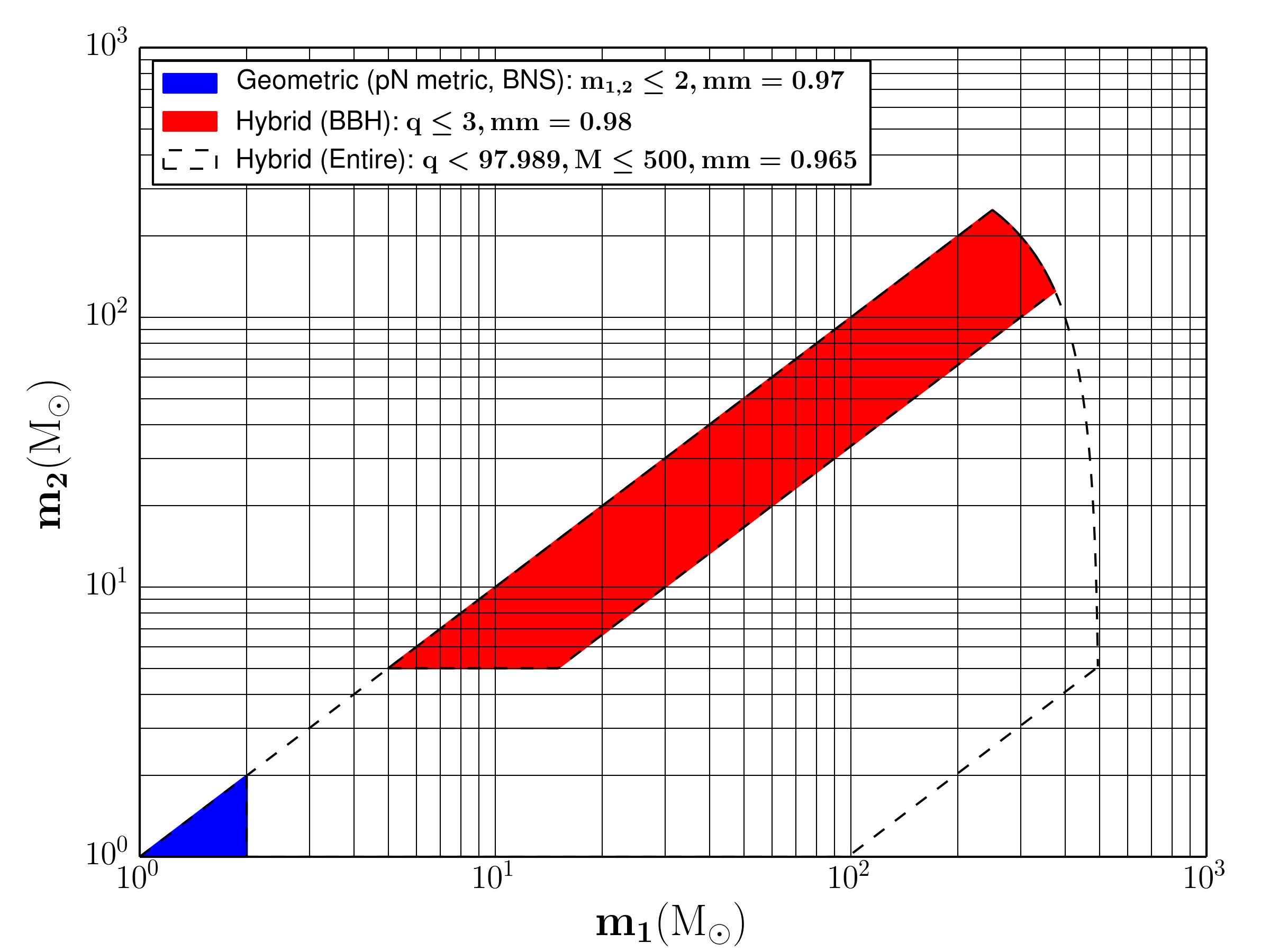}}}%
  \caption{Parameter space boundaries depicted in the component mass space for the O2 uber bank. The shaded regions in the left and right panels show the regions corresponding to the pre-computed seed banks used in the uber stochastic and hybrid banks respectively. The plot legends also indicate the method and the range of mass and spin parameters used in either method. Both the banks were generated using a optimized variable lower cutoff frequency $\flow$ and a reference frequency $\fref = 15\, \Hz$.}
  \label{fig:FullBank}
\end{figure*}

In order to understand the effect of these optimizations on the final bank, we have constructed toy hybrid template banks at a minimal match of $\Mmin = 0.98$ over nominal BBH search parameters where the total mass is in the range $M \in [10, 100]\, \msun$, component masses are in the range $m_1 \in [5, 500]\, \msun$ and $m_2 \in [5, 250] \msun$. The range of dimensionless spin magnitude was taken to upto $0.998$. The ER10 PSD was used in this experiment with a lower cutoff frequency at $\flow = 15 \, \Hz$ and $\fhigh = 1024\, \Hz$. The parameters corresponds to the red-shaded region in Fig.~\ref{fig:O2_uber_bank}. As we shall see later, such pre-computed banks are used to seed the O2 uber bank. These banks were compared against the vanilla and optimized stochastic banks constructed with identical parameters from random proposals in the $\{ \thetaz,\thetat, \xi_1, \xi_2 \}$ space. The results are summarised in Table~\ref{tab:BankEfficiency}. The optimizations explained earlier in this section have a profound effect on the final hybrid bank size. In this example, the optimized hybrid bank is found to have $22\%$ fewer templates as compared to the optimized stochastic bank. Fitting factor studies using $10\, 000$ injections indicate that the smaller hybrid bank provides marginally better coverage compared to the optimized stochastic bank.

Our main goal in this paper is to construct the optimized hybrid O2 uber bank which is described in the next section. We end this section with a few comments on the hybrid bank construction. By design, we can decouple the geometric and stochastic parts of the hybrid bank algorithm. To demonstrate this point, one may check that if we discard the shaded region in the flowchart, then the algorithm will choose one random proposal from $\rlist$ as a new template at a time and eliminate the points from $\rlist$ that lie inside the minimal match `sphere' centered at that point. This will continue until $\rlist$ becomes empty. This is exactly the stochastic placement turned on its head --- where the final bank is pared down from an initial list of a large number of random proposals, instead of being built from the ground up from single random proposals at a time (as described earlier). We call this the top-down approach of constructing a stochastic bank and have shown that it to be computationally more efficient than a bottom-up approach (where the parameter space metric is available) but result in the same bank sizes at a near-identical distribution of fitting factors~\cite{Roy-2017}. The reason of pointing out this ability to decouple the two parts is to highlight the fact that in regions of the parameter space that suffer from curvature and boundary effects, the geometric part of the algorithm is severely incapacitated and the construction essentially falls back to a vanilla stochastic placement algorithm. Thus, in the worst case scenario, the hybrid bank algorithm will perform at least as well as the stochastic placement algorithm.

\section{Constructing an optimized hybrid bank for Advanced LIGO}
\label{sec:FullBank}

In this section, we discuss the construction of an optimized hybrid template bank for the entire range of parameters in the O2 search and compare its performance against the optimized stochastic O2 uber bank. The search parameter space is summarized in Fig.~\ref{fig:O2_uber_bank} where the range of component masses are depicted. Component masses less than $2\,\msun$ are assumed to have dimensionless spin magnitude $\leq 0.05$. The spin magnitude of heavier component masses can extend upto $0.998$. In O2 search, the minimum length of the filtered waveform was set to be 150 ms to avoid the large number of triggers arising from short duration noise transients (glitches) in the data.




The PyCBC ``uber bank'' for O2 searches was constructed using a combined geometric-stochastic approach. First, the low-mass region of search space was covered by a pre-computed geometric seed bank at a minimal match 0.97 using the metric over post-Newtonian TaylorF2 waveforms~\cite{Brown-2012} with a fixed $\flow = 27\, \Hz$. This is depicted as the blue shaded region in Fig.~\ref{fig:O2_uber_bank} and the geometric seed-bank size has $129\, 000$ templates. Another seed bank over BBH parameters was also used which was constructed at a higher minimal match 0.98 as depicted by the red-shaded region in Fig.~\ref{fig:O2_uber_bank}. This extra coverage (with seed bank size of $20\, 070$ templates) was driven by the fact that the previously detected BBH sources were all found in this range of parameters. Finally, the optimized stochastic bank was generated by considering random proposals in the $\{\thetaz, \thetat, \xi_1, \xi_2 \}$ space at a minimal match of 0.965 by seeding these two banks. The BBH seed bank and final stochastic banks were generated by incorporating a variable lower cutoff frequency corresponding to $0.5\%$ acceptable loss in SNR as compared to a fixed lower cutoff frequency of $15\, \Hz$. 

The final stochastic O2 uber bank over the entire range of parameters space had $404\ 019$ templates.


A uber hybrid bank was constructed over the exact same range of parameters, seeded by pre-computed banks. The hybrid  bank construction was initialized using an initial set of $7 \times 10^6$ random proposals drawn from a uniform distribution in $\{\thetaz, \thetat, \thetats \}$ coordinates. The placement commenced from the center of the parameter space corresponding to component masses $m_{1,2} \equiv (10, 3) \msun$ and dimensionless spin components $\chi_{1,2} = 0$. Before commencing the construction over the full range of parameters, a geometric seed bank  consisting of $18\,517$ templates over the range of component masses corresponding to BNS systems was constructed at a minimal match 0.97 and the lower cutoff frequency fixed to $27\, \Hz$. A second pre-computed hybrid seed bank  consisting of $15\,671$ templates over BBH systems was also used. The final O2 optimized hybrid bank was found to have $350\,768$ templates.


Both these banks were generated using the harmonic mean of the ER10 PSDs of  H1 and L1 detectors. The reference frequency $\fref$ was set to $15\, \Hz$, below which there was no appreciable signal power for all the CBC templates within the search parameter space.

The reader may have noticed that BBH waveforms are used to cover a wide range of parameters corresponding to NSBH systems. GW signals from the late inspiral of NSBH binaries will contain higher order corrections due to the tidal deformation of the neutron star, depending on its equation of state, the mass ratio of the binary as well as the spins of the compact objects. However, the SNR of the post-inspiral part of NSBH binaries that Advanced LIGO may observe is likely to be small. Hence BBH waveforms are thought to provide a good approximation to such waveforms for the purpose of GW detection
In particular, NSBH waveforms with large mass ratios ($m_1/m_2 \gtrsim 6-9$) are expected to be identical to BBH waveforms~\cite{Shibata:2009cn}.


\begin{figure*}[htbp]
\centering
    \includegraphics[width=0.98\textwidth]{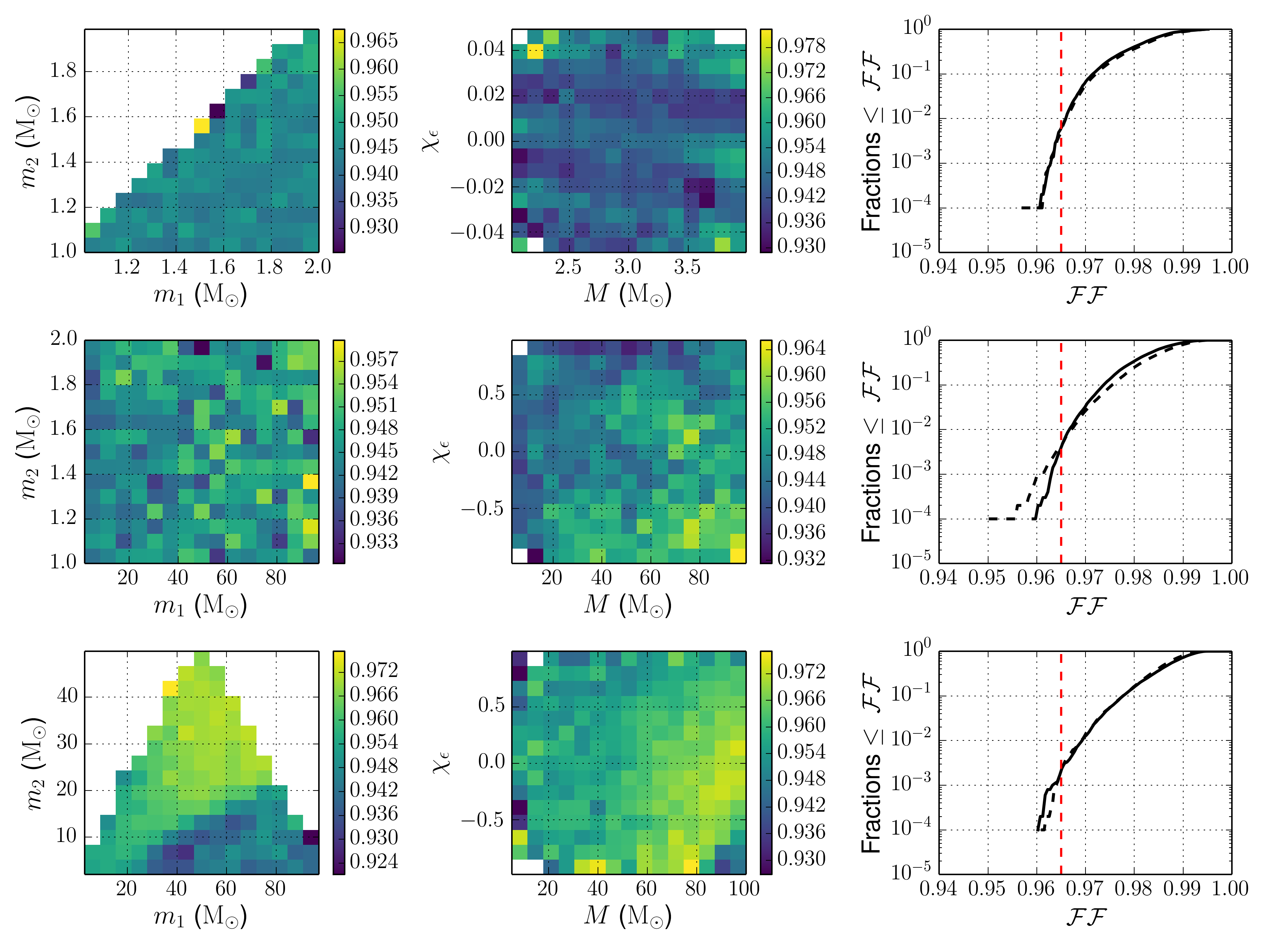}
  \caption{Performance of the banks over various regions of search space: BNS (\emph{top-row}), NSBH (\emph{middle-row}) and BBH (bottom-row) systems. Distributions of the fitting factors computed by injecting a set of $10\ 000$ $\seobnr$ signals in each of these three regions are depicted in the third column, where the \emph{solid line} and the \emph{solid dashed line} correspond to the hybrid and stochastic uber banks respectively. All the fitting factors are computed assuming ER10 harmonic PSD. A complete description of the performance of the banks are listed in Table~\ref{tab:O2bankSummary}. The heat maps in the first two columns show the average signal recovery fraction in each bin of the parameter space. It is evident that signal recovery fraction over the entire parameters space does not go below 0.924 ($>90$\%).}
  \label{fig:O2BankFF}
\end{figure*}

\subsection{Performance of the full banks}
\label{subsec:PerformanceO2Bank}


To quantify the bank coverage, we have carried out Monte-Carlo simulations to compute the distribution of fitting factors of the O2 uber banks against a set of $30,000$ injected aligned-spin $\seobnr$ signal waveforms with randomly chosen parameters. In this section, we present the results of this comparison and demonstrate that the both the optimized hybrid and the stochastic banks provide near identical coverage. The \texttt{pycbc\_banksim} program as implemented in \texttt{PyCBC}~\cite{pycbc} software was used for carrying out the simulation.

The \emph{fitting factor} $\mathcal{FF}$ of a template bank $\tlist$ with respect to an injected signal $h_{\ast}$ is defined as the maximum value of match over all the templates \cite{Apostolatos-1995}:
\begin{equation}\label{eq:ff}
  \mathcal{FF}(h_\ast) = \max_{\lambda \in \tlist} \text{\hspace{1mm}}  \match \left(\hat{h}_\ast, \hat{h}(\lambda) \right).
\end{equation}
Note that $1 - \mathcal{FF}(h_\ast)$ is the fractional loss of SNR in capturing the signal $h_\ast$ with the template bank $\tlist$. 
Here we assume that the signal model is the same for both templates and injections. As such, this loss arises from the discrete placement 
of the templates. It is desirable that the mismatch between the template and signal be bounded by a maximum value (usually set equal to 
$1 - \Mmin$), failing which can lead to an enhanced loss of detection rate. A bank is said to be \textit{effectual} if it can capture 
any arbitrary signals with a fractional loss in SNR less than this bound. The latter is proportional to $(1-\mathcal{FF})^3$ for a 
uniform distribution of source in the universe. For example, a bank constructed at a minimal match of $0.97$ corresponds to a maximum 
of $10\%$  loss in the detection rate. On the other hand, matched filtering searches using a densely packed template bank can increase 
the computational cost of the search. Thus, one requires a balance between computational cost and acceptable loss of detection rate. 
As noted elsewhere, the O2 uber banks (both stochastic as well as hybrid) were constructed at a $\Mmin = 0.965$ with the two seed 
banks constructed at $\Mmin = 0.97$ in the BNS region and at $\Mmin = 0.98$ in the BBH region.

The injected signals were generated from a uniform distribution over component masses and dimensionless spins. The range of the intrinsic parameters of injected signals (mass and spin) was taken to be identical to that of the O2 uber bank search parameters. These aligned-spin injections were classified in three regions of the search space: BNS systems with component masses between $[1-2]\msun$, BBH systems with component masses $m_{1,2} \geq 2$, and NSBH systems where the BH mass lies between $[2-100 ]\msun$ and NS mass lies in the range $[1-2] \msun$. 
The fitting factors for non-precessing binary systems are independent of the sky location, polarization angle and inclination angles. As such, the polarization angle was chosen from a uniform random distribution between $[0, \pi]$, the inclination angle was kept fixed corresponding to the face-on orientation of the binary system with respect to the line of sight and the sky location was taken from a uniform random distribution over the celestial sphere. 
The lower cutoff frequency for generating the injection waveforms was kept at a fixed $\flow = 15 \Hz$ while templates were generated at the optimum starting frequencies.


In the right most column of Fig.~\ref{fig:O2BankFF}, we show the results of the cumulative fitting factor distribution for hybrid and uber banks when both the template waveform and injected signals are modeled from $\seobnr$ approximant. The solid black line and the black dashed line represent the hybrid and uber banks respectively. The \textit{top right} plot of Fig.~\ref{fig:O2BankFF} corresponds to the injected BNS signals where, the cumulative fitting factor distribution shows that $0.6\%$ signals are recovered below a $0.965$ match for hybrid bank whereas $0.5\%$ signals are below this mark for the stochastic bank. The minimum fitting factor values for these set of injections are $0.96$ and $0.957$ in the hybrid and stochastic banks respectively. The right most plot seen in the second row of Fig.~\ref{fig:O2BankFF} corresponds to the full range of NSBH injections. In this plot, the cumulative fitting factors distribution for the hybrid bank shows that 0.39\% injections are found below 0.965 match with a minimum fitting factor of 0.96.  The black dashed line (corresponds to the stochastic bank) shows that 0.44\% injections are found below this mark with a minimum fitting factor 0.95. The performance of the banks against BBH injections is depicted in the \textit{bottom-right} plot Fig.~\ref{fig:O2BankFF} and shows that $0.23\%$ (hybrid bank) and $0.22\%$ (stochastic bank) of the injected signals are found below the mark $0.965$ match. Minimum fitting factor for both the banks were found to be at 0.96. From these three plots, it is evident that the hybrid uber O2 bank with $53 \,000$ fewer templates provides a near exact coverage as the O2 stochastic uber bank in all the three regions of the search space.

\begin{table*}
\centering
\begin{tabular}{ccccccc}
\toprule[1pt]
\toprule[1pt]
Binary system   \ \ \    &	Placement Algorithm \hspace{1mm} & Bank Size \hspace{1mm}  & \multicolumn{3}{c}{\% of $\mathcal{FF}<$ 0.965}     &   Comments \\
\cmidrule[0.8pt](lr{0.75em}){4-6}
& &  & BNS & BBH & NSBH & \\

 \midrule[0.8pt]
\multirow{2}{*}{\begin{tabular}[b]{@{}c@{}} BNS+BBH+NSBH \ \  \end{tabular}}
 &	Hybrid (with variable $\flow$) & $  350\, 768 $ &0.6 & 0.23 & 0.39     &  \multirow{2}{*}{\begin{tabular}[b]{@{}c@{}}  $\sim 14.5\%$ larger templtes\end{tabular}} \\ [0.5mm]
 &	Uber (with variable $\flow$)  & $ 404\, 019 $ & 0.5 & 0.22 &  0.44     &   \\ [0.5mm]
\bottomrule[1pt]
\bottomrule[1pt]
\end{tabular}
\caption{Summary of the O2 uber template banks. Hybrid banks were generated using a numerical $\seobnr$ metric in 3D dimensionless chirptime coordinates $\vec \theta \equiv \{ \thetaz, \theta_3, \theta_{3S}\}$ co-ordinates. On the other hand, the stochastic bank was constructed by considering random proposals in the four-dimensional $\{ \thetaz, \theta_3, \xi_1, \xi_2 \}$ parameter space using the exact match function. The complete description of the parameters for the full range of the search space is reported in Fig~\ref{fig:FullBank}. Please note that the O2 optimized hybrid bank took 40 hours over 50 CPUs while the O2 optimized stochastic uber bank was generated using 500 nodes over 25 days!}
\label{tab:O2bankSummary}
\end{table*}




As pointed out by Buonnanno et al.~\cite{Buonanno-2003},  the distribution of fitting factor may not be adequate to quantify the efficiency of a given template bank against short duration signal injections with less signal power. Therefore to get a complete picture, one also should compute the \emph{signal recovery fraction} (SRF) of the injected signals against a template bank. The SRF is defined as:
\begin{equation}
\label{eq:srf}
  \mathrm{SRF} = \frac{\sum_{i}\mathcal{FF}^3(h^i) \ \sigma^3(h^i)}{\sum_{i}\sigma^3(h^i)},
\end{equation}
where, $\sigma(h^i)$ is the horizon distance and is proportional to the signal power. The horizon distance is defined as the farthest luminosity distance of an optimally inclined (face-on) and oriented (overhead) CBC source at a fixed canonical SNR.



In the first and second columns of Fig.~\ref{fig:O2BankFF} we show the SRF for injected signals in different combination of parameters. 
The three different rows correspond to the three mass ranges of the search space corresponding to BNS, NSBH and BBH systems respectively. We have computed the average SRF in each bin over mass-spin parameters by using Eq.~\ref{eq:srf}. 
It is expected that a bank must be constructed to achieve an SRF greater than 90\% over the entire search space. From these three rows of plots in  Fig.~\ref{fig:O2BankFF}, it is clearly evident that SRF does not fall below 92.4\% over the entire search space, thereby validating these banks.

\section{Summary and conclusions}
\label{sec:conclusions}

We report the construction of effectual template banks for GW searches from compact binaries in data from future advanced LIGO runs, after incorporating several new optimizations to the geometric-random \emph{hybrid} placement algorithm proposed by us in an  earlier paper~\cite{Roy-2017}. These hybrid template banks are constructed in an effective 3 dimensional parameter space using the semi-numerical metric over the parameter space of aligned spin CBC waveforms. The algorithm combines the space-efficiency of a geometrical lattice of points along with the robustness of the stochastic placement algorithm using which, one can place a more efficient bank in parameter spaces having slowly varying metric. No additional fine tuning is needed for accommodating curvature and edge effects while placing the templates. The new optimizations include the use of a variable lower-cutoff frequency and imposing a lower bound on the template duration which not only improve the reach of LIGO searches to high-mass binary black hole systems up to several hundred solar masses but also keeps a check on the overall bank size. We also capitalise on a degree of freedom wherein the underlying geometrical lattice of template points is suitably oriented as determined by the enclosing boundary of the target search space. We also describe a computational efficient way of including exact match values within the hybrid placement algorithm. The combined effect of the new optimizations presented in this paper leads to a $\sim 34\%$ reduction in the hybrid bank size as shown in Table~\ref{tab:BankEfficiency}.

The hybrid algorithm assumes local flat patches that extend upto a spherical region of radius $2 \dmax$ for relatively high values of $\Mmin = 0.97$. For parameter spaces  where the metric varies even more slowly,  it may be possible to extend the notion of such a flat patch up to twice of this radius. This may result in faster template placement with additional reduction in the overall bank size without affecting the coverage of the bank significantly. Finally, the random proposals in $\rlist$ may be sprayed more efficiently over the dimensionless chirp-time coordinates according to the spatial density determined by $\sqrt{|\mathbf{g}|}$. We would like to explore these ideas in future.

We demonstrate the efficacy of the optimized hybrid algorithm by constructing toy seed banks over a nominal BBH range of parameters. As seen in Table~\ref{tab:BankEfficiency}, such banks are about $22\%$ smaller in size as compared to the optimized stochastic bank constructed over identical parameter space. We also constructed the O2 uber hybrid bank and compared it in detail with the one used in LIGO's O2 search. The latter was generated ~\cite{Canton-2017} using pre-computed seed template banks in the BBH and low-mass region by an optimized version of the stochastic  algorithm with a variable starting frequency. The distribution of fitting factors and signal recovery fraction obtained from Monte-Carlo signal injections over a wide range of parameters establish the fact that the hybrid bank, with $53\, 000$ fewer templates, can provide nearly identical coverage. 

The LIGO Scientific Collaboration is gearing up for the O3 science run expected to start in March 2019. In view of this, we have generated uber O3 hybrid banks with identical parameters as those for the O2 uber bank and have used the ER13 harmonic PSD. The corresponding bank found to have $412\, 000$ templates. This implies that by using the optimized hybrid bank algorithm one should be able to analyze the data from a more sensitive O3 search using nearly the same computational resources as the previously concluded O2 offline search. On the other hand, assuming the O3 target PSD, we end up with a bank with $1.6$ million templates. In O3 searches, it may be possible to increase the parameter space boundary to include systems with total mass more than $500\, \msun$ due to improvements in the low-frequency sensitivity of the detectors. 


While the effective banks presented in this paper are in three-dimensional space, one may have to consider placing template banks in higher dimensional spaces for future CBC searches in LIGO data using binary waveforms with fully precessing spins or with higher-order modes. In view of this, we have also sketched an outline for extending the hybrid formalism to arbitrary number of parameter-space dimensions by the use of $\anstar{n}$ lattices in Appendix~\ref{sec:HybridInHigherDimensions}. 

Our effectual template banks are ready to be used in CBC searches in upcoming observation runs of advanced LIGO and Virgo detectors.

\appendix

\section{Hybrid bank construction in higher dimension}
\label{sec:HybridInHigherDimensions}

In a previous work, we proposed the hybrid template placement algorithm in a  3-dimensional parameter space by using the $\anstar{3}$ (or BCC) lattice. In this paper, we propose new optimizations which leads to further improvements. The algorithm can be easily extended to higher dimensional parameter spaces with weakly varying metric by the use of $\anstar{n}$ lattice (where $n\geq3$) in the geometrical part of the algorithm. Such banks may be needed in future CBC searches using template  waveforms with fully precessing spins. 4D stochastic banks have been constructed for CBC searches using non-spinning template waveforms with higher order modes~\cite{Harry-2017}. Such banks have been shown to increase the sensitivity of searches for high mass ratio and high total mass systems. It would be prudent to use the hybrid method, such as the one presented in this work in such cases.


To illustrate the performance of the hybrid algorithm and also to compare it against the stochastic method in higher dimensions, we generate both hybrid and stochastic banks in 3 and 4 dimensions using a constant metric $g_{ij} \equiv \delta_{ij}$ for which the minimal match spheres have unit radius by construction. Here, $\delta$ refers to the Kronecker delta function. These templates are placed inside a spherical volume of radius $10$ in the corresponding dimensions. These choices for the metric (constant) and volume (sphere has the smallest surface to volume ratio) were made to theoretically minimize the curvature and boundary effects. 
The hybrid banks were found to have $35\%$ fewer templates as compared to the stochastic bank in three-dimension and $25\%$ fewer in four-dimension. As these toy banks have been constructed in idealized conditions free of curvature effects and minimum boundary effects, these numbers may as well serve as theoretical limits to the improvement in bank sizes from a hybrid construction in 3 and 4 dimensions respectively. 

\begin{figure}[htbp]
\centering
  \includegraphics[width=0.475\textwidth]{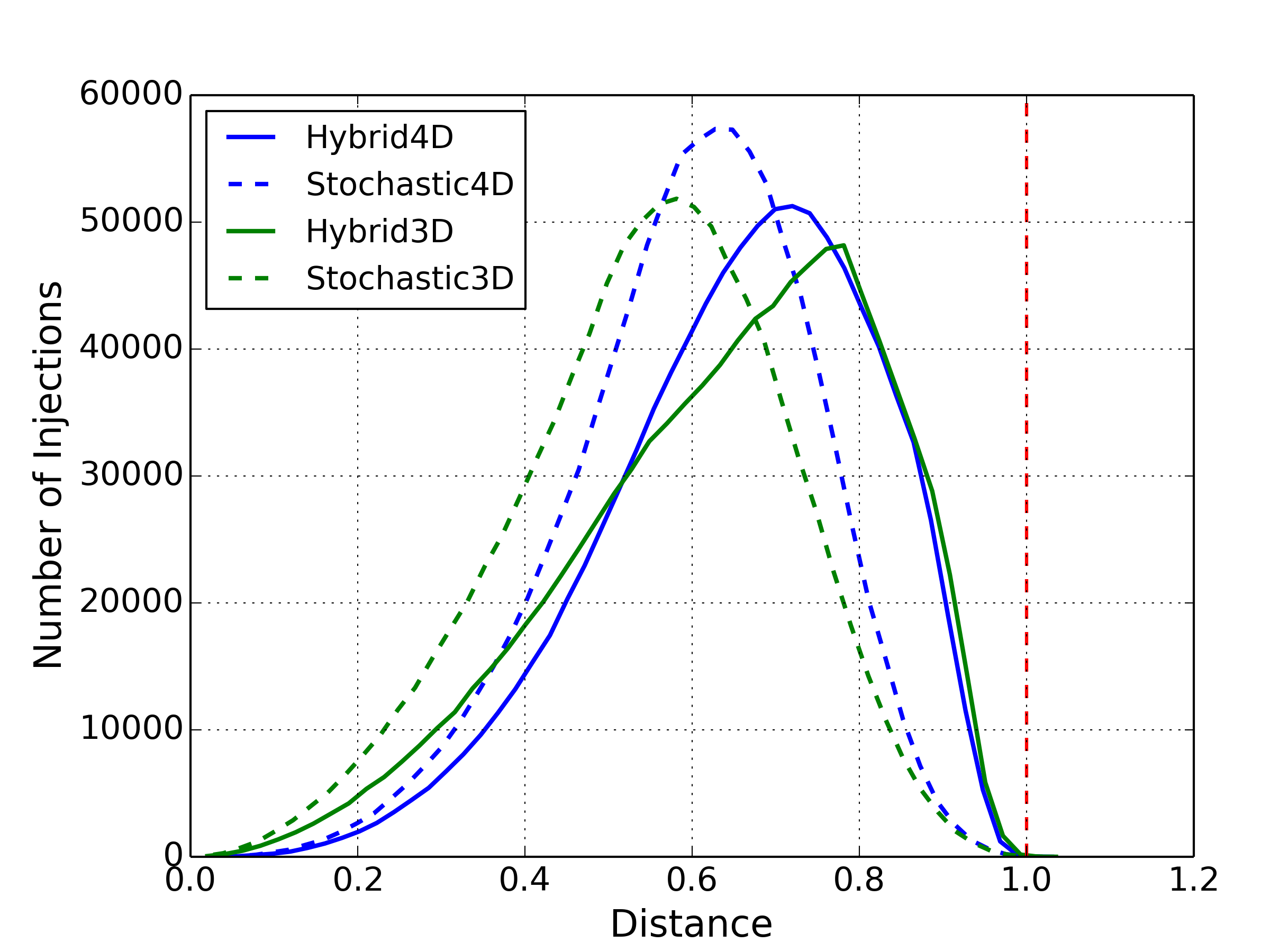}
  \caption{Histogram of distances between a set of $10^6$ uniformly distributed random points and templates for $n=3,4$ dimensions. The template banks were generated using both the stochastic and hybrid method by assuming a constant metric $g_{ij} = \delta_{ij}$ (corresponding to unit covering radius) within a spherical volume of radius = 10 so as to minimize the effects of curvature and boundary effects. Such injection studies have also been carried out for fully lattice based template banks~\cite{Wette-2014}.}
  \label{fig:bankstest}
\end{figure}

The banks were tested for coverage by injecting $10^6$ random points and calculating the Euclidean distance to the nearest template point which is related to the mismatch through Eq.~\ref{eq:MisMatch}. A histogram of such distances is shown in Fig.~\ref{fig:bankstest}: for both 3-D and 4-D cases. The peak of the distribution is seen to shift to the right for the hybrid banks in both cases which allude to the over-density of points in the stochastic banks as expected. For the given constant metric, it would interesting to check how close the hybrid algorithm is to generating a fully lattice-placed template bank~\cite{Wette-2014} in the bulk. Near the vicinity of the parameter space boundaries, the interaction between lattice and random template placement will result in imperfections.

Any $n$-dimensional lattice can be constructed by the orthogonal projection of the $n+1$ dimensional cubic lattice $\mathbb{Z}^{n+1}$ onto the hyperplane perpendicular to a suitably chosen vector $\vec{v} \in \mathbb{Z}^{n+1}$ \cite{Sloane2011}. In particular, choosing $\vec v = \vec 1$  (column vectors of 1s) leads to the construction of the $\anstar{n}$ lattice ~\cite{McKilliam-2008} using the projection matrix $\mathbf{G}$ defined as:
\begin{equation}
\label{eq:AnStarProjection}
 \mathbf{G} = \mathbf{I}- \frac{\vec{1}^{T}\vec{1}}{n+1},
\end{equation}
where $\mathbf{I}$ is the identity matrix.
It can be shown that the $\anstar{n}$ lattice provides optimal sphere covering where each sphere of radius $r = \sqrt{|\mathbf{G} \mathbf{G}^{T}|}$ is placed at the lattice points. The nearest neighbours of any point in this lattice can be identified to be the location of lattice points within the distance $2r$. We can transpose this idea for finding the $\anstar{n}$ nearest neighbours for the template placement problem by scaling $r$ to the covering radius $\dmax$ of each template, which is related to the minimal match of the bank through the expression $\dmax = \sqrt{1 - \Mmin}$.


\vspace{6pt}

\begin{acknowledgments}
A.~S. would like to thank LS and B Sathyaprakash for useful discussions - in particular, for the suggestion on numerical calculation of the metric. We especially thank Ian Harry and Tito Dal Canton for their help and several suggestions in carrying out this work. We acknowledge useful discussions with Collin Capano, Nathaniel Indik, Nathan K Johnson-McDaniel and Chad Hanna for useful discussions at different stages of this work. S.~R. thanks IIT Gandhinagar for research fellowship and fellow graduate students (Amit Reza, Chakresh Singh, Md.~Yousuf) for useful discussions and Toral Gupta and Niladri Naskar for help with the manuscript. S.~R. and A.~S. thank ICTS-TIFR for hospitality and support, where the initial phase of this work was carried out. P.~A.'s research was supported by a Ramanujan Fellowship from the Science and Engineering Research Board (SERB), India and by the Max Planck Society through a Max Planck Partner Group at ICTS-TIFR.  A.~S.'s research was partly supported through the SERB grant EMR/2016/007593 from Department of Science and Technology.
\end{acknowledgments}

\bibliography{reference}{}
\bibliographystyle{apsrev4-1}

\end{document}